# Identification of new gold lines in the 350 – 1000 nm spectral region using laser produced plasmas

M. Charlwood[1], S. Chaurasia[2*], M. McCann[2], C. Ballance[2], D. Riley[1] and F. P. Keenan[2]

[1]Centre for Light-Matter Interactions, School of Mathematics and Physics, Queen's University Belfast, University Road, Belfast, BT7 1NN, UK

[2]Astrophysics Research Centre, School of Mathematics and Physics, Queen's University Belfast, University Road, Belfast, BT7 1NN, UK

Correspondence should be addressed to S. Chaurasia; s.chaurasia@qub.ac.uk

## Abstract

We present results from a pilot study, using a laser-produced plasma, to identify new lines in the 350 – 1000 nm spectral region for the r-process element gold (Au), of relevance to studies of neutron star mergers. This was achieved via optical-IR spectroscopy of a laser-produced Au plasma, with an Au target of high purity (99.95 %) and a low vacuum pressure to remove any air contamination from the experimental spectra. Our data were recorded with a spectrometer of 750 mm focal length and 1200 lines mm$^{-1}$ grating, yielding a resolution of 0.04 nm. We find 54 lines not previously identified and which are not due to the impurities (principally copper (Cu) and silver (Ag)) in our Au sample. Of these 54 lines, we provisionally match 21 strong transitions to theoretical results from collisional-radiative models that include energy levels derived from atomic structure calculations up to the 6s level. Some of the remaining 33 unidentified lines in our spectra are also strong and may be due to transitions involving energy levels which are higher-lying than those in our plasma models. Nevertheless, our experiments demonstrate that laser-produced plasmas are well suited to the identification of transitions in r-process elements, with the method applicable to spectra ranging from UV to IR wavelengths.

**Keywords:** Neutron Star Mergers; LIBS; Line Identification; Atomic Data; Atomic Spectroscopy

**Introduction:**

Gravitational waves were first detected arising from a binary black hole merger by (Abbott et al. [1]). Subsequently, such waves were also observed from a neutron star merger (NSM), GW170817. The electromagnetic counterpart to GW170817, resulting from the material ejected during the merger, was identified by Smartt et al. [2] as AT2017gfo and is termed a kilonova. Such sources are luminous at optical and IR wavelengths, hence allowing the detailed study of NSMs, which can help address many important topics in astrophysics, such as the expansion rate of the Universe (Schutz 1986 [3]).

Of particular relevance to the present paper is the role of NSMs in the production of heavy elements via rapid neutron capture (r-process). The r-process pathway is responsible for creating 50 % of all the heavy elements ($Z > 30$) and is the only source of elements beyond bismuth at $Z = 83$ (Thielemann et al. [4]). Several sites for the production of r-processed elements have been theorised, including NSMs but also collapsars and magnetorotational supernovae (see Cowan et al. [5] for a full list and discussion of possible r-process formation sites). However, NSMs are the only astrophysical sources where the production of r-process elements is confirmed, and also appear to be responsible for most of their creation (Cowan et al. [5]). Observations of the expanding material from an NSM – i.e., kilonovae – hence provide the only possibility to detect and quantify the r-process in situ. This in turn will allow some of the biggest open questions in astrophysics to be addressed, namely how and where are the heavy elements from Fe to U made?

Unfortunately, the large outflow velocities of kilonovae ejecta (~ 0.1–0.3c) leads to spectral lines being both Doppler broadened and shifted from their rest wavelengths (see e.g., Gillanders et al. [6]). Consequently, individual lines cannot be resolved and employed to determine element abundances. Instead, researchers need to use radiative transfer codes such as TARDIS (Kerzendorf et al. [7]) to model the spectra and assess the presence (or otherwise) of r-process elements (Gillanders et al. [6]). Two of the most interesting elements to search for evidence of their presence are platinum (Pt) and gold (Au), as the former is one of the most abundant 3$^{rd}$ peak r-process elements predicted to be created in NSMs, while Au should also be abundant (Goriely et al. [8]).

However, the atomic data required as an input to kilonovae models for these elements is very limited, including for line wavelengths. These are important, as any absorption or emission due to such features may hence be missed by models if the lines are not included. One can use theoretical wavelengths (e.g., Gillanders et al. [6]), but these must be matched and corrected to experimental measurements, to ensure that transitions are properly identified and that their (calculated) transition probabilities (which scale as $\Delta E^3$ for allowed lines) can be corrected for the energy difference between theory and experiment.

Very limited line identification data available for Pt as illustrated by the NIST Atomic Spectra Database (Kramida et al. [9]), which lists only 27 lines of Pt I, 3 of Pt II and zero for Pt III at wavelengths > 800 nm, where Gillanders et al. [6] note that one might expect to find strong, observable spectral features. In the case of Au, Bromley et al. [10] obtained experimental data for Au I and Au II, covering 187–800 nm. They employed Au-tipped probes, constructed by electroplating steel cylinders with a Ni layer followed by an Au one, which were inserted into current-carrying discharges on the Compact Toroidal Hybrid (CTH) device at Auburn University. In total they detected 94 new Au I and Au II lines, with only 8 lying above 700 nm. The steel and Ni components of the probes, plus other gases, resulted in the presence of many additional lines, with Bromley et al. [10] noting that the 187–800 nm region contains thousands of Fe I and Fe II transitions alone. Consequently, a good number of Au lines were contaminated by possible overlapping, although Bromley et al. [10] obtained plasma discharges with Au absent to allow this to be assessed and minimised. The lack of atomic data at longer wavelengths for Pt and Au is also illustrated by the TARDIS plasma models shown in Gillanders et al. [6], where no absorption nor emission features are predicted beyond 1000 nm since very few transitions are included in TARDIS that can produce any observable deviation from the continuum.

Given the above discussion, we have initiated an ambitious programme to obtain accurate experimental line wavelengths for 3$^{rd}$ peak r-process species important for NSM studies, focusing on the unexplored spectral region above 800 nm and in particular, in the longer term, the infrared range (1–5 μm) relevant to kilonovae observations. For our programme we will use laser-generated plasmas of high purity samples in high vacuum, greatly reducing contamination and hence blending with other elements. Here we present some results from a pilot study on Au, in the optical-NIR range, to demonstrate the validity and effectiveness of our approach.

**Experimental setup:**

A schematic of our experimental setup is shown in Figure 1, where a Q-switched Nd:YAG laser (GCR Class IV, 1064 nm wavelength), with energy and pulse duration of up to 1.2 J and 10 ns, respectively, was focused onto the target. The laser can be operated in single shot mode as well as 10 Hz mode. Our primary target was a high purity (99.95 %), 0.50 $\pm$ 0.05 mm thick solid Au sample, with principal impurities of Cu (500 ppm) and Ag (300 ppm), all other impurities being lower than 15 ppm. To assess any possible blending from these contaminants, we also obtained data with high purity (both 99.99 %) Cu and Ag using the same experimental setup. In addition, some shots were undertaken for 99.95 % purity Pt and 99.99 % purity palladium (Pd), with the latter being a minor contaminant for Pt (50 ppm). Known lines for both elements were also employed to help determine the wavelength calibration for the spectrometer (see below).

The Au target was placed in a vacuum chamber on a motorised X-Y-Z translation stage and pumped down to $10^{-5}$ mbar to remove air contaminates and hence generate a purer Au plasma. In the data presented here, the laser energy was set to 400 mJ and the focal spot diameter fixed to either 3 mm or 1.3 mm, which resulted in laser intensities on target of $6\times10^{8}$ $Wcm^{-2}$ and $3\times10^{9}$ $Wcm^{-2}$, respectively. Spectra were obtained over the 350–1000 nm wavelength region using a Shamrock 750 spectrometer coupled to an iKon-M 1024 × 1024-pixel CCD camera. For the 350–700 nm range, the experiment was performed at both of the above laser intensities. However, for the NIR region (700–1000 nm), the quantum efficiency of the CCD drops (~10 % at 1000 nm) and the larger spot size was used which helped to increase the plasma volume and yield a better signal. As indicated in Figure 1, the plasma plume was imaged onto the spectrometer, and in particular, the region ~1 mm from the solid surface was imaged onto the 50 μm slit**.** This allowed a lower density region with limited Stark broadening to be sampled. Our procedure for estimating the plasma temperature and density is discussed in the data analysis section.

To remove the contribution of background scattered light to the experimental spectra, the target assembly was covered with a black box with only two openings, one for the input laser pulse and the other at 90 ° to this for viewing by the spectrometer. For longer wavelength (700–1000 nm) measurements, a 700 nm long-pass filter was fitted to block shorter wavelength lines from appearing in higher order. A 1064 nm cut-off filter was used for the whole experiment to block the laser beam contribution. We employed a 1200 lines $mm^{-1}$ grating for our experiment, yielding an RMS value of resolution of 0.02 nm for more than 355 data points (the manufacturer specification for this value is 0.04 nm). However, using this grating meant that each measurement covered only a ~11 nm section of the Au spectrum on the CCD camera. Hence to obtain data over the whole 350–1000 nm range of interest, the grating angle needed to be changed ~70 times. Each spectrum was the sum of 150 laser shots, to increase the signal-to-noise and facilitate the identification of weak emission features. For each set of 150 shots, the position of the shots on the target was changed, and before each set the target surface was cleaned using 10 – 15 shots.

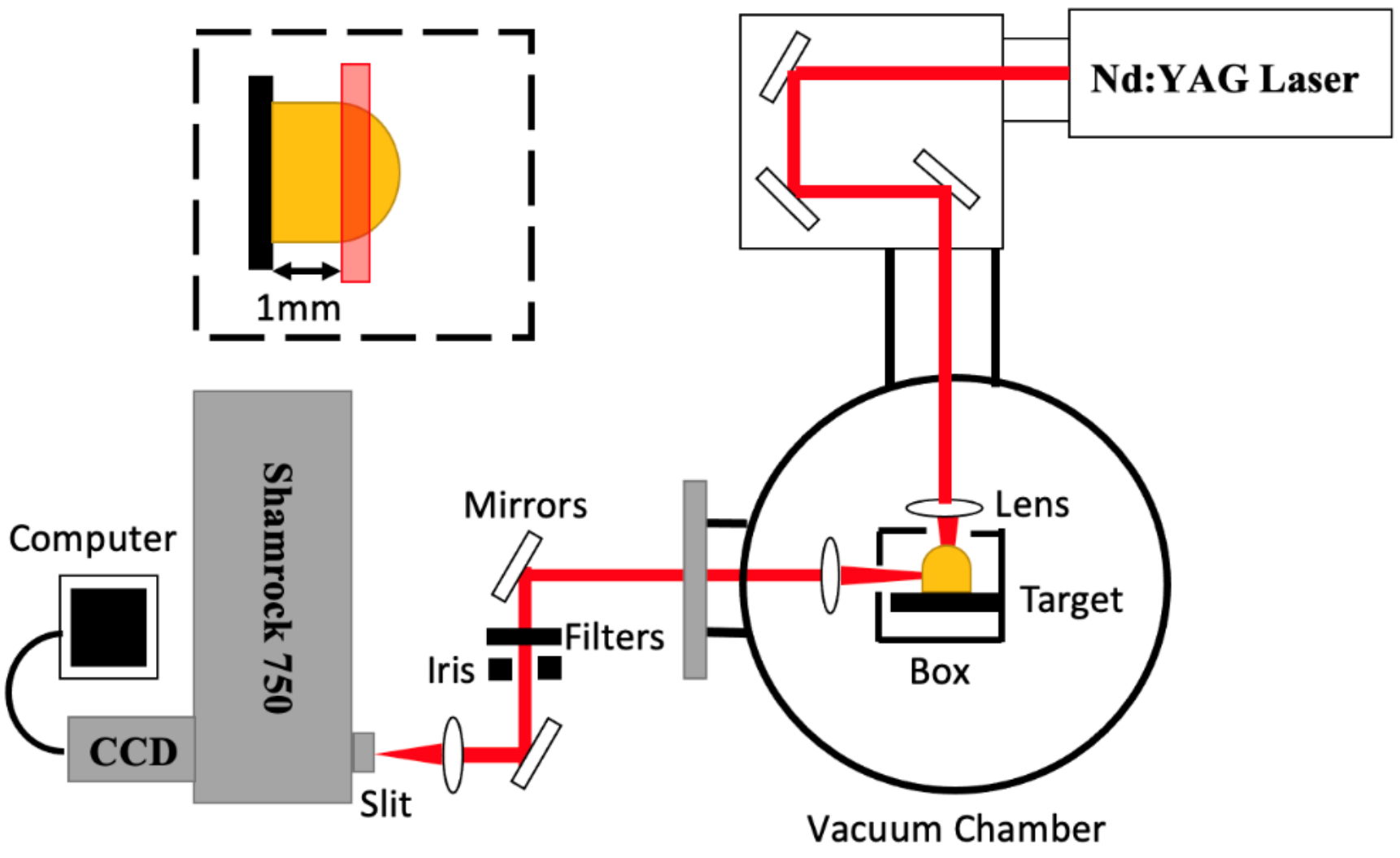

**Figure 1:** Top-view schematic of the experimental setup, with the Au target placed inside the vacuum chamber. An Andor Shamrock 750 spectrometer was coupled to an Andor iKon-M CCD camera which collected the Au spectral data. The inset shows the section of the Au plasma that was focused onto the slit of the spectrometer outside the vacuum chamber.

Wavelength calibration of the spectrometer was performed with a mercury Hg I lamp using 6 well-known strong Hg lines from the NIST database covering the 300 – 600 nm range. Spectrometer software controlled the grating and allowed the measurement of wavelengths for spectral features. The resulting calibration showed good agreement with our measurements over this wavelength range, with a maximum difference of $\Delta\lambda$ = 0.04 nm, as shown in Table 1.

**Table 1:** Difference $\Delta\lambda$ between our spectrometer Hg line wavelengths and NIST values

| NIST value (nm) | Spectrometer (nm) | $\Delta$ (nm) |
|---|---|---|
| 365.02 | 365.03 | -0.01 |
| 404.66 | 404.65 | 0.01 |
| 435.83 | 435.84 | -0.01 |
| 546.07 | 546.08 | -0.01 |
| 576.96 | 576.92 | 0.04 |
| 579.07 | 579.03 | 0.04 |

In addition, lines of Au, Ag, Pt, Pd and Cu covering 300 – 1000 nm with previously determined experimental wavelengths from NIST showed good agreement with our measured values, to better than 0.04 nm. In Figure 2 we plot the differences between our experimental values, as measured with our spectrometer, and those from NIST. As we can see, the largest differences are ~ 0.04 nm. The root means square (RMS) deviation using all 355 observed lines was 0.02 nm, better than the resolution specified by manufacturer (0.04 nm). The known lines from NIST for Au, Ag, Pt, Pd and Cu that were used for wavelength calibration are given in Annexure A.

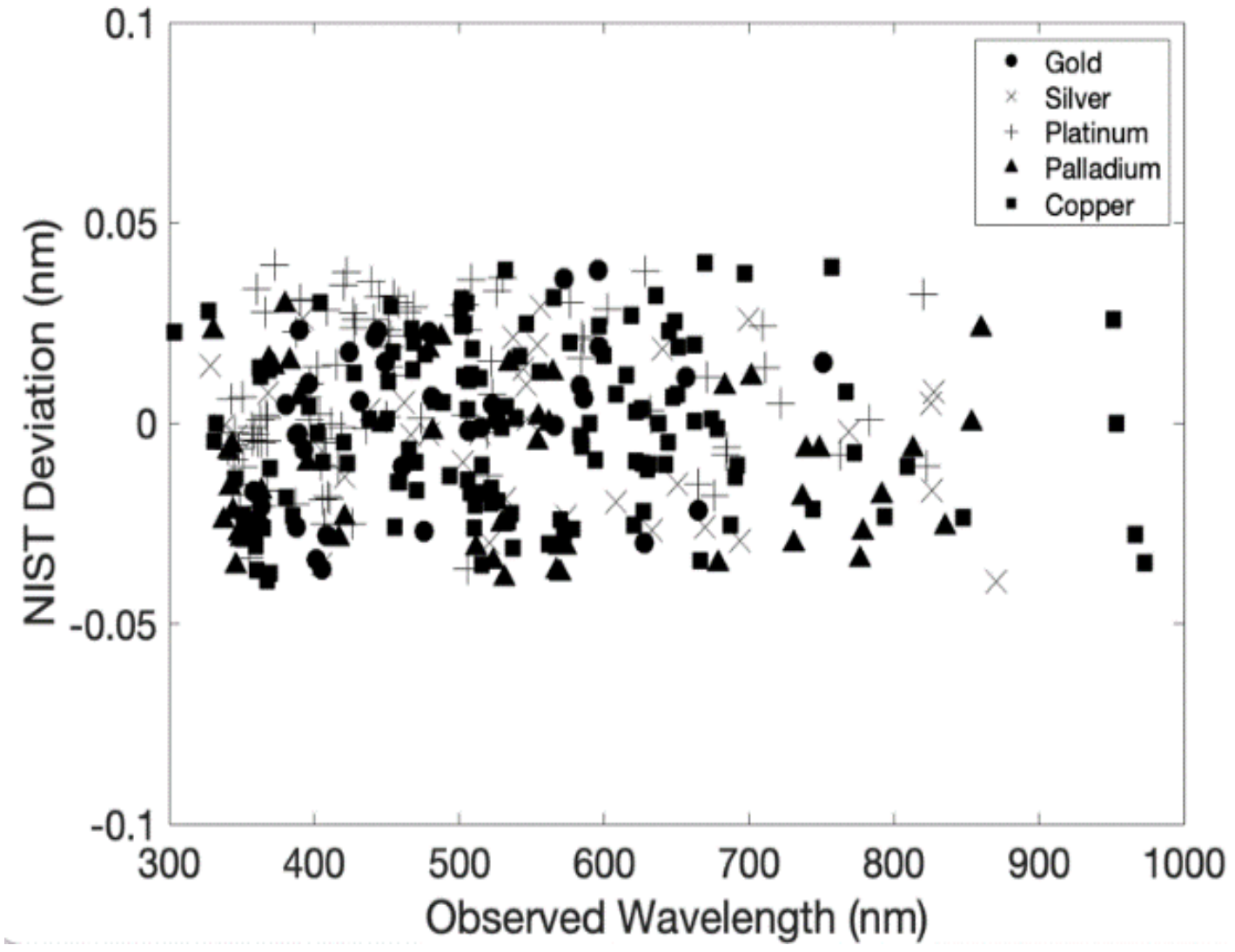

**Figure 2:** Plot of the differences between tabulated NIST wavelengths and our spectrometer measurements for relevant elements.

**Data analysis:**

The main motivation of our experiment was to search for new Au lines, particularly in the relatively unexplored 700 – 1000 nm wavelength region. However, for completeness, we extended our work to the 350 – 700 nm range. In Figure 3 we show experimental Au spectra covering 350 – 700 nm obtained at both laser spot intensities. As noted above, only 11 nm of spectrum is measured at a time, and the results in Figure 3 are hence a composite of many spectra, each the result of merging 150 laser shot. The signal is larger for the lower laser intensity shots (larger focal spot of 3 mm), as a result of the greater plasma volume created. We note that the transmission of the BK7 glass used in the vacuum chamber window (10 mm) falls-off rapidly at wavelengths below 400 nm, hence the small signal levels for the 350 – 400 nm data, and our lack of line detections below 350 nm. In Figure 4, lines at wavelengths beyond 700 nm are clearly detected, which as discussed in Section 5 are not listed by NIST nor reported by Bromley et al. [10] and in earlier work.

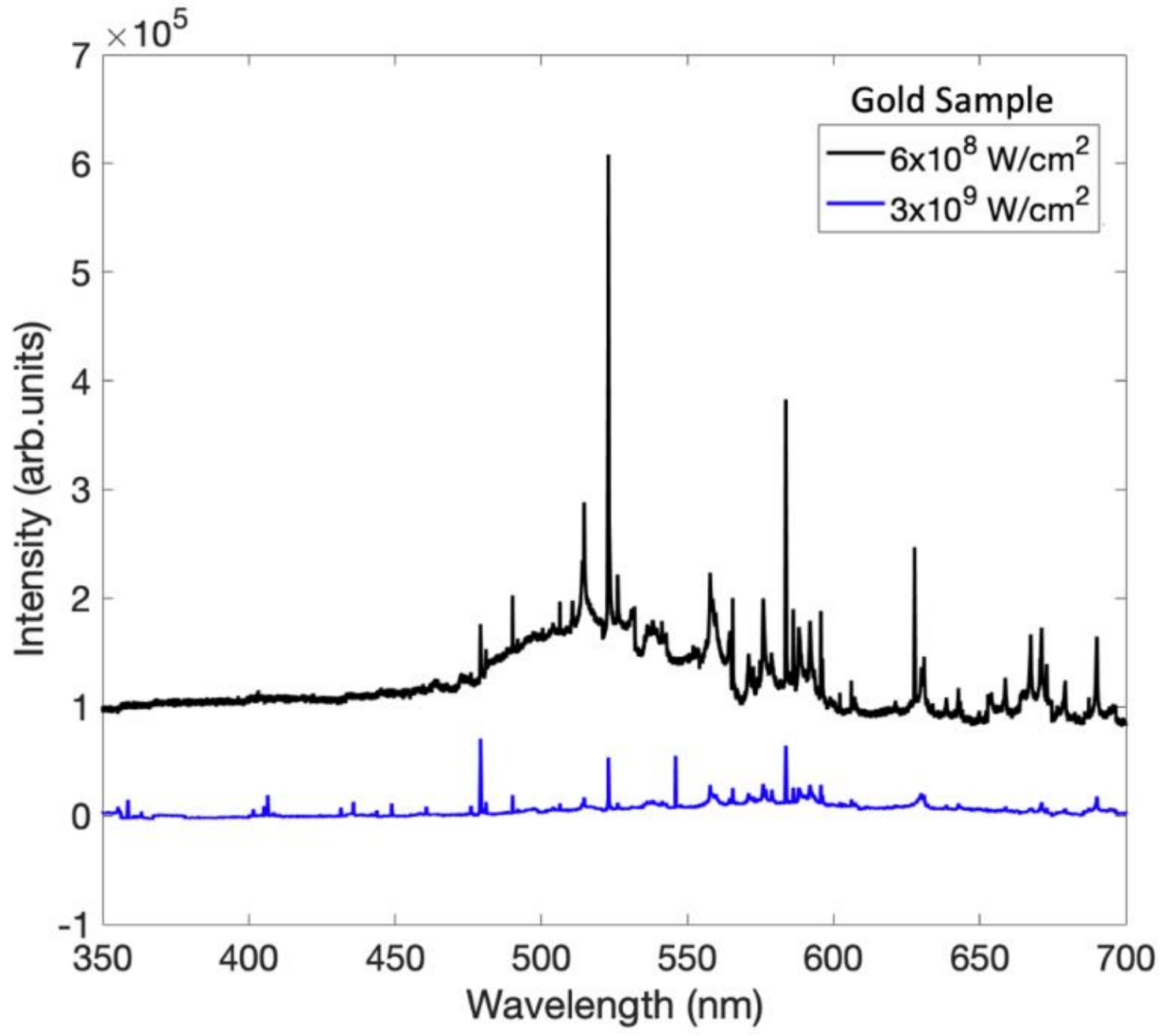

**Figure 3:** Composite Au spectra covering 350 – 700 nm recorded at two different laser intensities. The focal spot diameters of the black and red spectra were 3 mm and 1.3 mm, respectively, for a fixed laser energy of 400 mJ. The vacuum chamber window is 10 mm BK7 glass and so transmission starts to drop at wavelength below 400 nm. Spectra have been vertically displaced for ease of viewing.

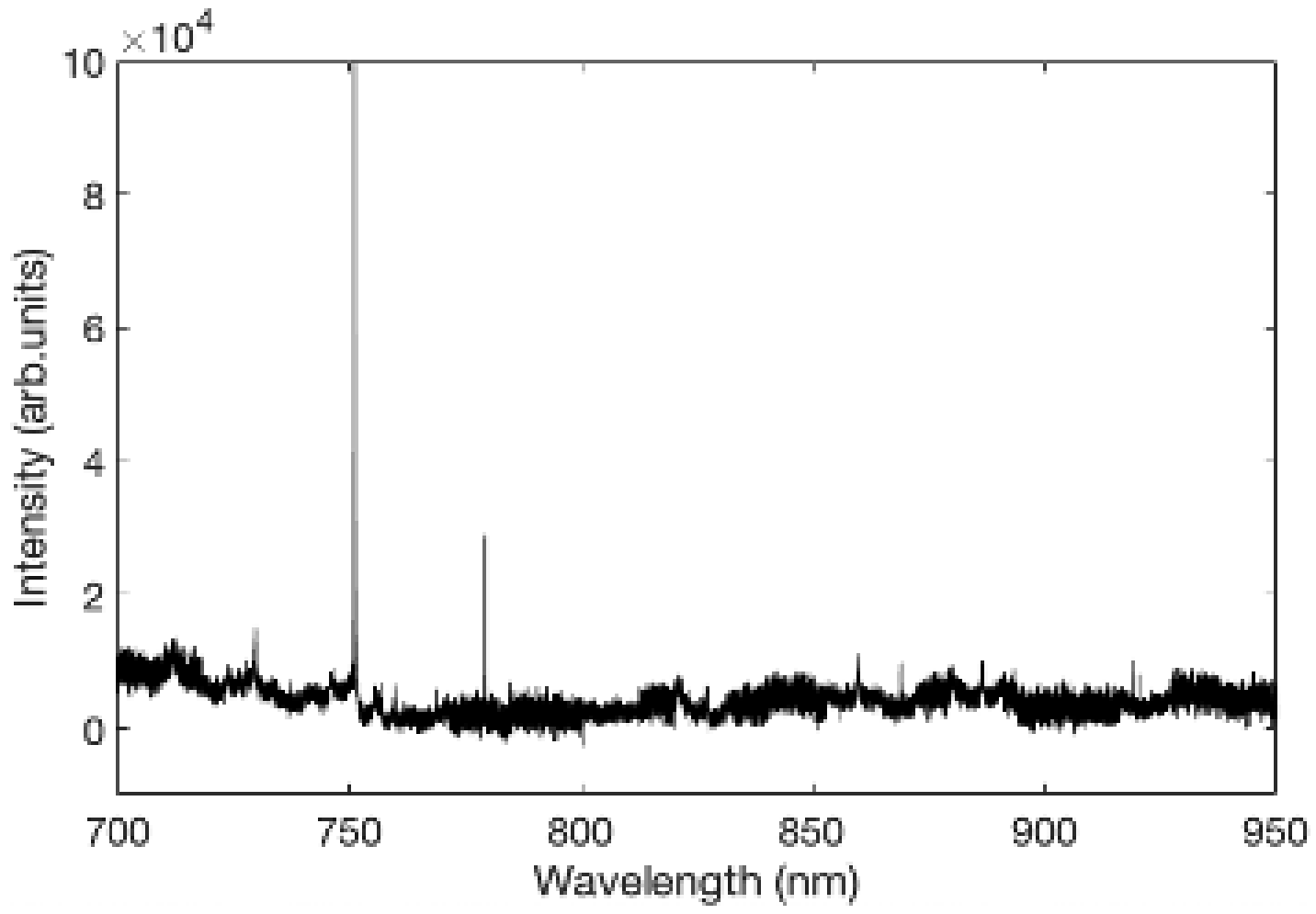

**Figure 4:** Composite Au spectrum obtained for the 700 – 950 nm range at a laser intensity on target of $6\times10^{8}$ $Wcm^{-2}$. The quantum efficiency of our CCD camera limited the measurements to < 1000 nm.

A previous study of Au in the 180–800 nm region was undertaken by Rosberg and Wyart [11] using a hollow cathode in helium (He) and neon (Ne) background gas environments. As noted earlier, Bromley et al. [10] reported some new Au identifications in the same spectral range, although in their work there are also lines from several impurities and background gases which contaminate the spectral region of interest. By contrast, an advantage of our method is that relatively pure Au samples can be used under vacuum to minimise spectral contamination. Nevertheless, our samples do have some minor impurities, and hence it is important we determine which lines in our spectra might be present from elements other than Au. We have therefore obtained spectra for Cu (purity 99.99 %) and Ag (purity 99.99 %) samples procured from Goodfellow. Considering the maximum contamination of 500 ppm and 300 ppm from these elements, respectively, in our Au targets, only their strongest lines are likely to be important.

In Figure 5 we compare the experimental spectra for Au, Ag and Cu. All the spectra were taken at the same fixed distance from the target surface, and the focal spot was also kept constant, so that any geometry effects should be minimal. Most of the lines in the Au data did not match those in the Cu and Ag spectra. There are some instances where there is a match in wavelength, to within our resolution, of Cu or Ag lines with those from Au. However, in these cases the lines were generally assessed not to be due to Cu or Ag, as their intensities were inconsistent with their impurity levels. Given this, we believe these lines are additional unidentified Au transitions.

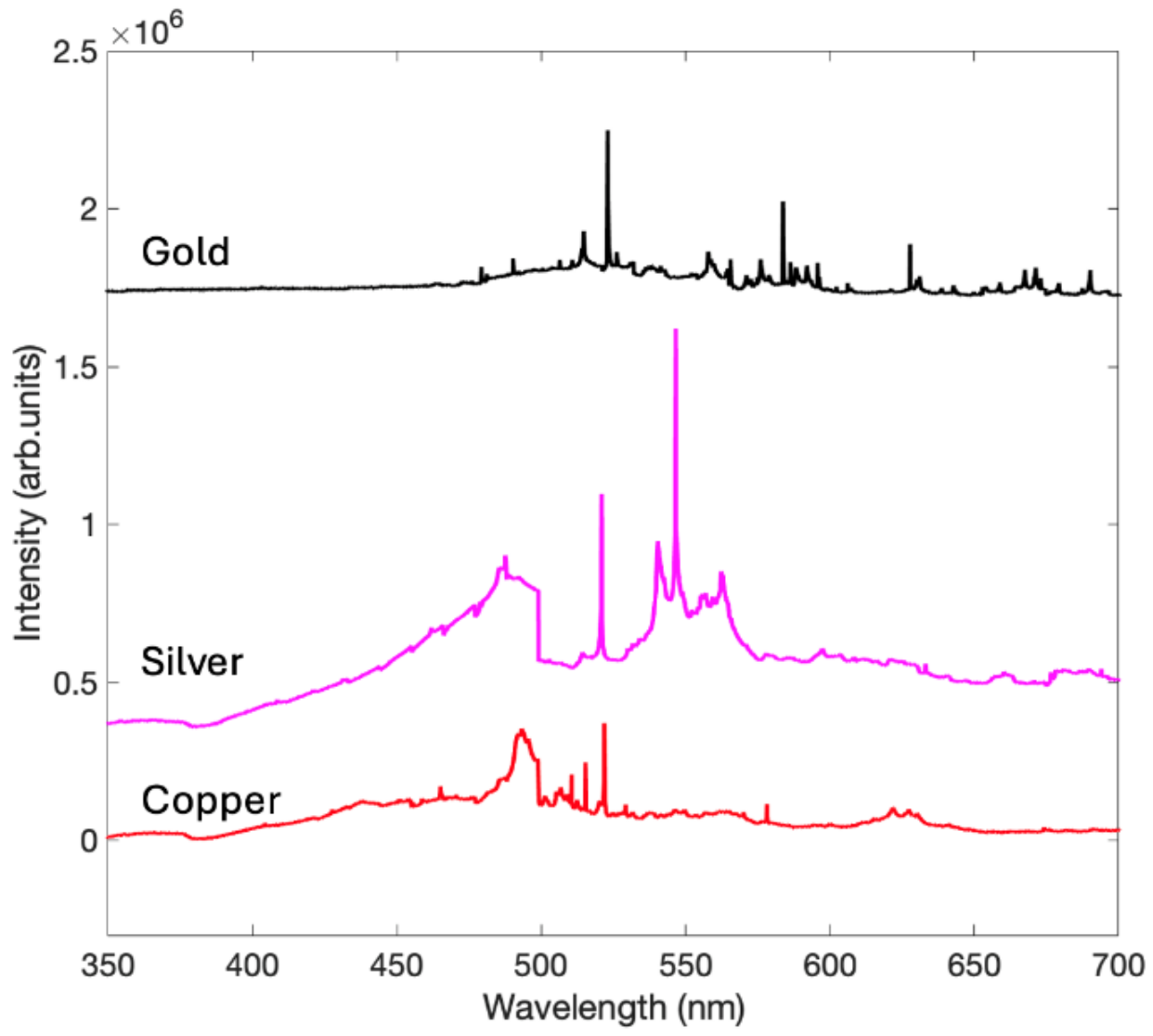

**Figure 5:** Experimental spectra of Au (99.95 % purity), Cu and Ag targets in the 350 – 700 nm range, obtained at a laser intensity of $6\times10^{8}$ $Wcm^{-2}$. Spectra have been vertically displaced for ease of viewing.

**Estimation of plasma temperature and density:**

The electron temperature and density of the plasma for the two experimental laser intensities were calculated, as these would be useful when comparing our line identification results with the predictions of atomic structure codes. With the assumption that the plasma is in local thermodynamic equilibrium (LTE), the population of excited states, from the same ion stage, provides an estimate of the electron temperature $T_e$ using the Boltzmann plot method [12-13] given by

$$ln\left(\frac{I\lambda}{g_j A}\right) = \frac{-1}{k_B T_e} E_j + ln\left(\frac{4\pi Z}{hcN_0}\right) \quad (1)$$

where $I$ is the spectral line intensity, $\lambda$ the line wavelength, $A$ the transition probability, $g_j$ the energy level statistical weight, $E_j$ the upper state energy level, $Z$ the atomic number of Au, $N_0$ the total species population and $k_B$ Boltzmann's constant. In Table 2 we list the Au I line data from the NIST database used for this analysis. Boltzmann plots for laser intensities of $6\times10^{8}$ $Wcm^{-2}$ and $3\times10^{9}$ $Wcm^{-2}$ are shown in Figure 6, and from these the corresponding values of the electron temperature are 0.5 ± 0.15 eV and 0.85 ± 0.21 eV, respectively.

**Table 2:** Spectroscopic Au I data used to estimate the plasma electron temperature.

| Wavelength (nm) | Energy of Lower State, $E_i$ (eV) | Energy of Upper State, $E_j$ (eV) | Statistical Weight, $g_j$ | Transition Rate, ($A/10^8$ $s^{-1}$) |
|---|---|---|---|---|
| 406.51 | 4.63 | 7.68 | 4 | 0.85 |
| 479.27 | 5.11 | 7.69 | 6 | 0.89 |
| 481.17 | 5.11 | 7.68 | 4 | 0.16 |
| 506.46 | 2.66 | 5.11 | 4 | 0.01 |
| 583.73 | 4.63 | 6.76 | 2 | 0.30 |
| 627.82 | 2.66 | 4.63 | 2 | 0.03 |
| 751.07 | 5.11 | 6.76 | 2 | 0.42 |

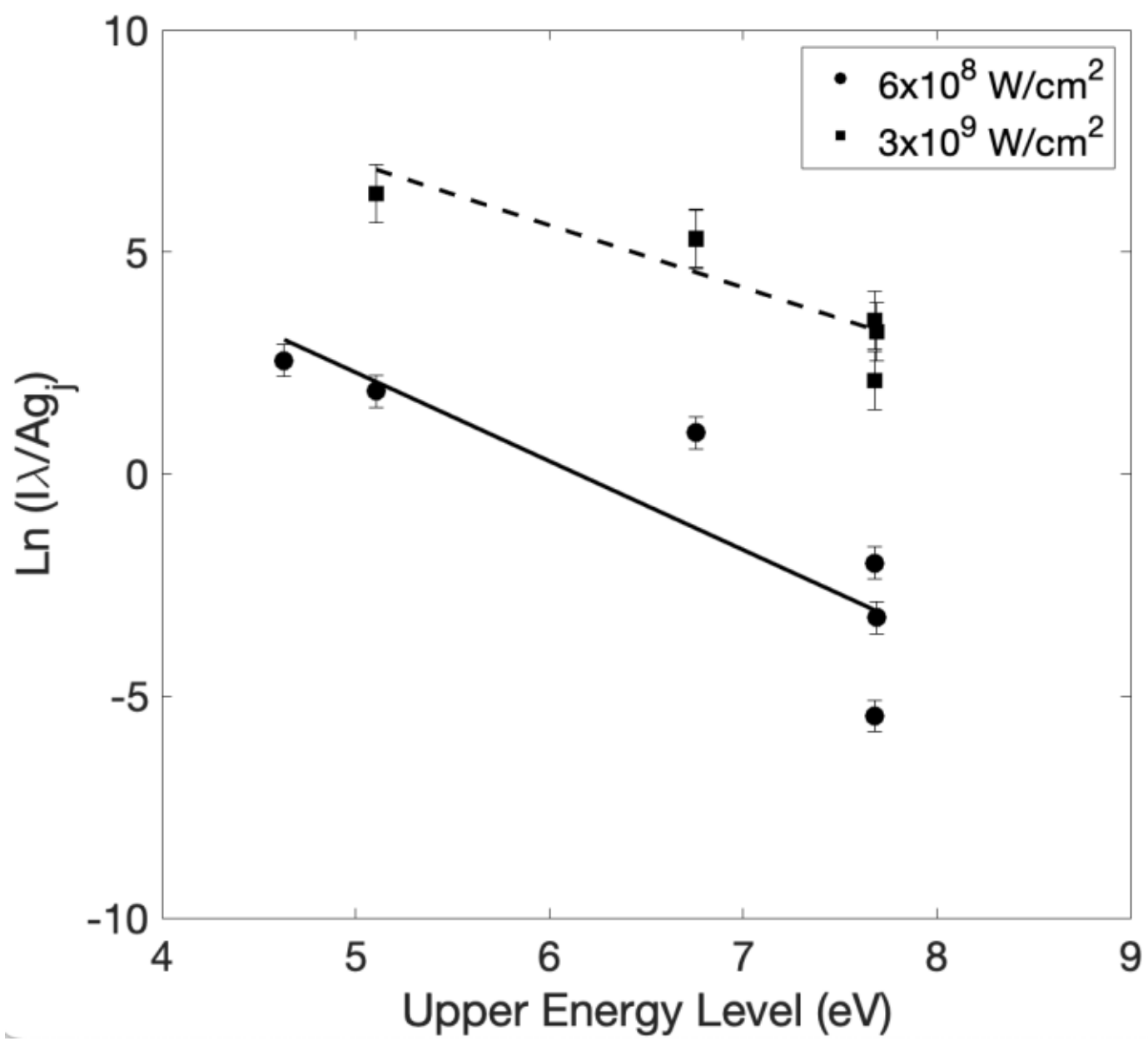

**Figure 6:** Plot of ln ($I\lambda/g_jA$) as a function of upper state energy level for the Au I emission lines in Table 2, at two laser intensities under vacuum. The $3\times10^9$ $Wcm^{-2}$ data have been scaled by a factor of 3 for clarity.

For our assumption of LTE to hold, it is required that the atomic and ionic states are populated and depopulated predominantly by electron collisions rather than by radiation, meaning the electron density must be sufficiently large to ensure a high collisional rate. The electron density required to achieve LTE in a plasma is given by the McWhirter criterion [14-15]:

$$N_e > 1.6 \times 10^{12} T_e^{\frac{1}{2}} \Delta E^3 \tag{2}$$

where $\Delta E$ is the largest energy difference between the upper and lower-level transitions, which in our experiments is for the Au I 406.51 nm line, with $\Delta E$ = 3.05 eV. Using this value of $\Delta E$ and the electron temperatures derived above, the electron densities required for LTE are calculated to be $3.5\times10^{15}$ cm$^{-3}$ and $4.5\times10^{15}$ cm$^{-3}$ for the $6\times10^{8}$ Wcm$^{-2}$ and $3\times10^{9}$ Wcm$^{-2}$ laser intensities, respectively.

To confirm the prediction made in Equation 2, we have generated theoretical populations for the two levels involved in the Au I 406.51 nm line, using the collisional radiative modelling code ColRadPy (Johnson et al. [16]). In Figure 7 the density dependence of the populations is shown, to demonstrate when the levels enter the LTE regime. At both temperatures of 0.5 ± 0.15 eV and 0.85 ± 0.21eV, the level populations are in LTE above the predicted densities of $3.5\times10^{15}$ cm$^{-3}$ and $4.5\times10^{15}$ cm$^{-3}$.

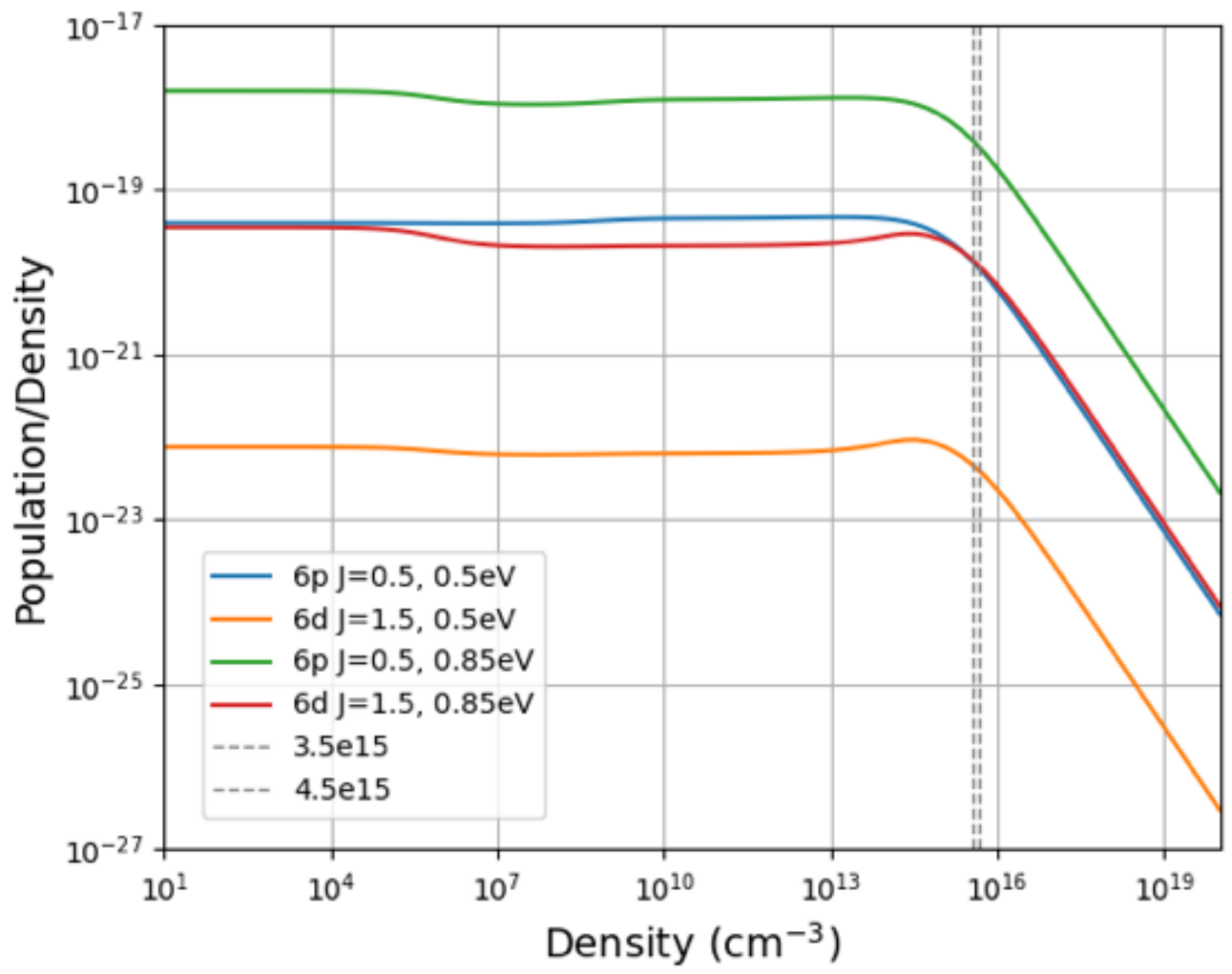

**Figure 7:** Density dependence of the populations of the lower and upper energy levels of the Au I 406.51 nm line. The behaviour of the populations above the predicted densities of $3.5\times10^{15}$ cm$^{-3}$ and $4.5\times10^{15}$ cm$^{-3}$ indicate that the levels are in LTE.

Stark broadening of the optically thin spectral lines was used to calculate the electron densities $N_e$ of the Au plasmas for the two laser intensities, using the equation [17-18]:

$$\Delta\lambda_{1/2} = 2W\left(\frac{N_e}{10^{16}}\right) \tag{3}$$

where $\Delta\lambda_{1/2}$ is the FWHM of a Stark broadened line and W is the electron density impact width parameter, which is weakly dependent on the electron temperature through the equation:

$$W(T_e) = 4.8767 \times 10^{-4} + 1.6385 \times 10^{-8} T_e - 1.8473 \times 10^{-13} {T_e}^2 \tag{4}$$

Values of $W$ were determined from Equation 4, while $\Delta\lambda_{1/2}$ was measured through line fitting using a Lorentzian function, as shown in Figure 8 for the Au I 506.46 nm line, taking into account the instrumental broadening of 0.02 nm. For the $6{\times}10^8$ $Wcm^{-2}$ and $3{\times}10^9$ $Wcm^{-2}$ laser intensities, the FWHM of this Stark broadened line are 0.062 nm and 0.076 nm, respectively. Using Equations 3 and 4, the electron densities were then calculated to be $5.4{\times}10^{17}$ $cm^{-3}$ and $6{\times}10^{17}$ $cm^{-3}$, respectively. These values are much higher than those determined from the McWhirter criterion, indicating that the Au plasma is in LTE and hence our earlier assumption of this was correct.

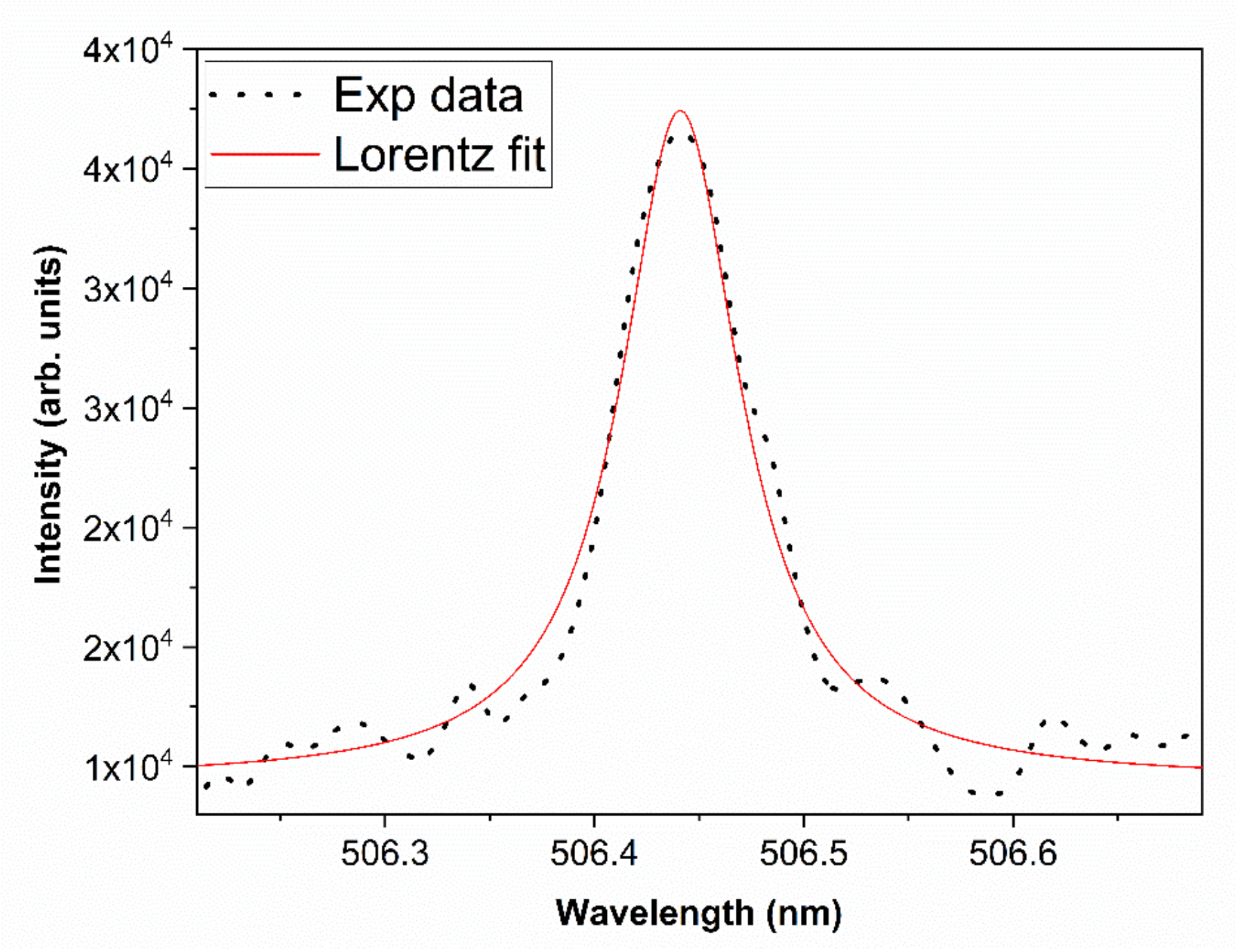

**Figure 8:** Lorentzian fit of the experimental Au I 506.46 nm line with a laser intensity of $6{\times}10^8$ $Wcm^{-2}$.

**Atomic physics and line intensity calculations:**

Synthetic spectra for Au I and Au II were generated using atomic data from the GRASP$^0$ (General-purpose Relativistic Atomic Structure Package) [19] and a parallelised version of DARC (Dirac Atomic R-matrix Code) [20]. GRASP$^0$ solves the coupled radial form of the Dirac equation to provide electron orbitals, which are subsequently employed in the R-matrix electron-impact excitation code DARC to calculate collision strengths. These then provide electron-impact excitation/de-excitation rates for subsequent collisional-radiative modelling.

Details of the Au I and Au II atomic structure, and electron-impact excitation calculations for the former, are given in McCann et al. [21] For the first two ion stages of Au, there are extensive experimental energy levels and identified transitions given within the NIST database, though these are not exhaustive as the current paper shows. The focus of the Gillanders et al. [6] and McCann et al. [21] papers was to aid in the determination of possible low-lying transitions observed under NSM plasma conditions. Therefore, in pursuit of greater spectroscopic accuracy, low-lying levels of Au I and Au II were shifted to NIST energy level values where available. The shifted energies included the first 26 levels in Au I and 44 levels in Au II and are shown in Tables 3 and 4.

Both sets of electron-impact excitation calculations for Au I and Au II were subsequently convolved with a Maxwellian for a range of temperatures to provide effective collision rates. The NIST adopted energy levels, calculated E1, M1, E2 and M2 Einstein A-coefficients and the Maxwellian averaged rates were then used to construct a collisional-radiative matrix that was solved at the experimentally determined electron temperature of 0.85 eV and electron density of $6\times10^{17}$ cm$^{-3}$. This matrix was solved within a quasi-static approximation for which, in each ion stage, the excited state populations are weighted by the ground state population. The consequence of this is that the absolute heights of the synthetic spectra may have less meaning than the relative heights used in line intensity ratios. Synthetic spectra for Au I, Au II and Au III were plotted with the experimental data to aid the line matching. To illustrate this, a spectrum containing the synthetic and experimental data, covering the 400 – 500 nm region, is shown in Figure 9a and b.

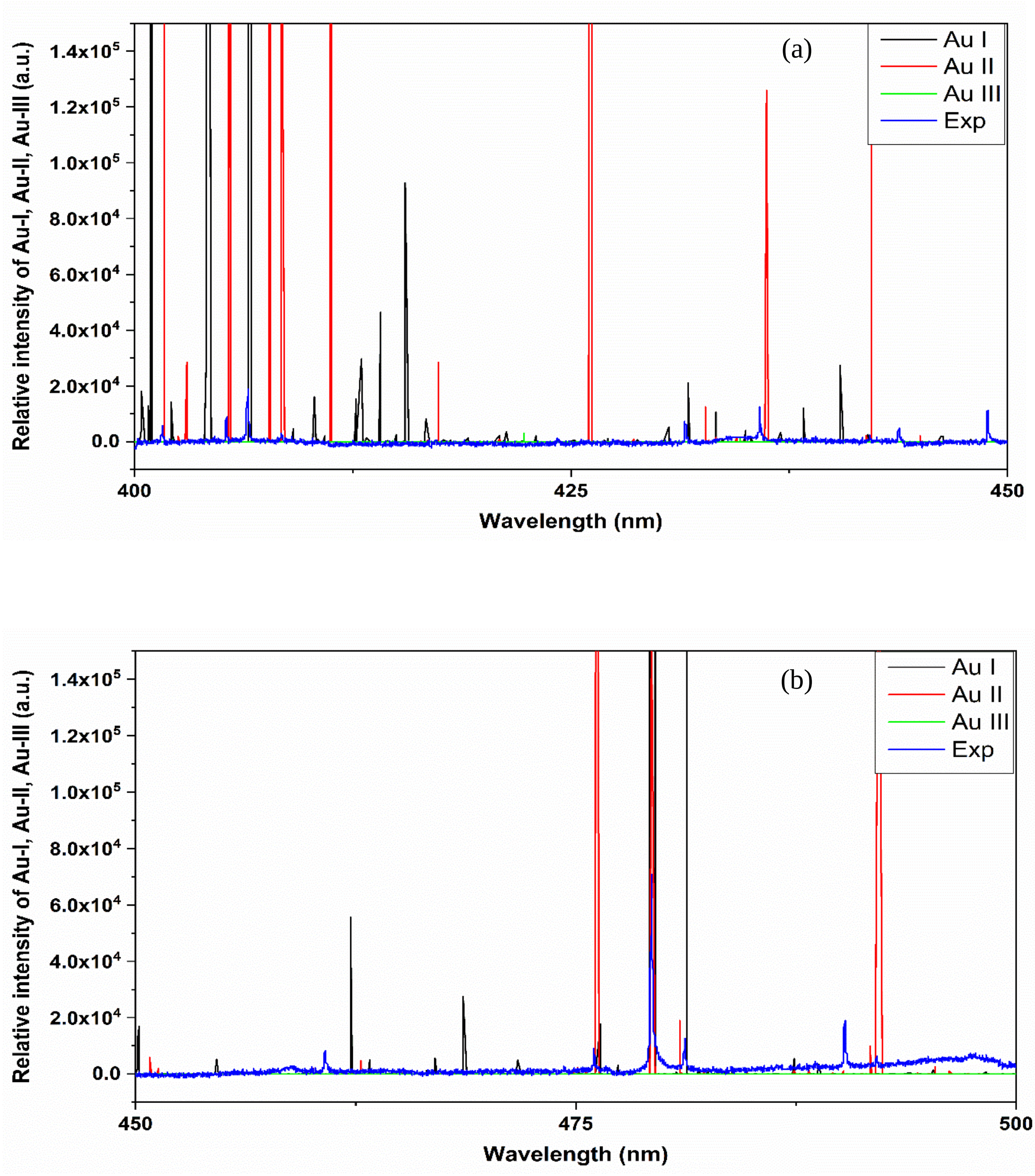

**Figure 9:** Comparison of experimental data with the synthetic spectra of Au I, Au II and Au III in the (a) 400 – 450 nm and (b) 450 – 500 nm regions. Intensities of lines in the synthetic spectra have been scaled by a factor of $10^4$ to better show weaker transitions, and also to facilitate having experimental and synthetic data in the same graph. Experimental spectra were recorded at a laser intensity of $3\times10^9$ Wcm$^{-2}$.

**Results and discussion:**

Our work is similar in some respects to that of Bromley et al. [10], but there are a number of key differences. Firstly, Bromley et al. [10] investigated the 197–764 nm region while we cover 350–1000 nm. Secondly, the Bromley et al. [10] experiments were at a temperature of 10 eV and electron density of $10^{12}$ $cm^{-3}$, and most of the lines they identify lie shortward of 400 nm with only a few longward, which is to be expected as the UV transitions will be stronger at higher temperatures. In our experiment, the temperature is 0.85 ± 0.21 eV and the density is $10^{17}$ $cm^{-3}$, and hence we have only measured 9 lines between 350 – 400 nm. Thirdly, Bromley et al. [10] compared their results to data primarily from two references, namely Platt & Sawyer [22] (PS) and Ehrhardt & Davis [23] (ED), both of which contain many of the Au lines present in the NIST database. However, we compare our data with Bromley et al. [10], PS [22], ED [23] and Rosberg & Wyart [11] (RW). PS [22] recorded new energy levels and lines of Au I and Au II using a hollow cathode discharge in He between 60–1000 nm and calculated a < 0.01 nm wavelength uncertainty within the vacuum region, comparable to ours. They reported 2000 new lines and energy levels, and NIST includes many of these in their database. ED [23] also used a hollow cathode lamp with a pair of electrodes to produce an Au spectrum to study Au I –III and reported Au I lines between 264–925 nm, and Au II lines from 93 – 476 nm. However, they had neon and argon impurity features in their spectra due to these filler gases being present in their assembly, and hence removed the corresponding lines from their results. Most if not all of the Au lines and energy levels listed in the NIST database are from ED [23].

In the 700 –1000 nm range of particular interest in our work, NIST (ED [23]) lists 3 lines with 1 matching our experiment (751.07 nm). However, in this wavelength region we find 20 lines not in NIST, but PS [22] lists 7 lines which match our experimental measurements and atomic physics calculations for Au II.

A comparison of the Au I and Au II lines detected in our experimental spectra with the NIST database (PS [22], ED [23], RW [11]), Bromley et al. [10]), plus our atomic physics calculations, is presented in Tables 5 and 6. For those identified lines where all three sets of data (experiment, NIST/Bromley et al. [10] and calculations) are comparable, there is good agreement. This indicates that the atomic physics calculations for both Au I and Au II are valid, and can be employed to identify the Au lines we have found in our experimental data that are not listed in NIST nor Bromley et al. [10].

In Tables 5 and 6, 'unshifted' indicates the theoretical Au energy levels are not shifted to the those in NIST as they are absent from the latter. However, due to their intensity distribution and proximity to our experimental wavelengths, they have been provisionally matched to our data. If an experimental line is matched with the calculations and is not listed as unshifted, then the theoretical energy levels have been shifted to those in NIST, and hence should be more accurate. For Au I, 26 out of 200 energy levels were shifted to the NIST values (ED [23], Brown & Ginter [24] (BG), Moore [25] (M)), and 44 out of 431 for Au II (RW [11], M [25]). In some cases, there is a small deviation of 0.06 nm in the line wavelengths, greater than the 0.04 nm resolution limit of the experiment, even for the shifted energy levels with the NIST database. This is due to the fact that, in the NIST database, the wavelength of a transition may be taken from one reference and the relevant energy levels from others. For example, $5d^{10}6p\ ^2P_{1/2}$ (37359 $cm^{-1}$) and $5d^{10}6d\ ^2D_{3/2}$ (61952 $cm^{-1}$) are listed as the lower and upper energy levels, respectively, for the 406.51 nm Au I line. However, the $5d^{10}6p\ ^2P_{1/2}$ (37359 $cm^{-1}$) level is from ED [23] while $5d^{10}6d\ ^2D_{3/2}$ (61952 $cm^{-1}$) is from M [25]. Similarly for Au II, the 679.7 nm line predicted by the atomic physics calculations indicates $5p^65d^97s\ ^1D_2$ (108631 $cm^{-1}$) and $5p^65d^97p\ ^3F_3$ (123348 $cm^{-1}$) from NIST as the lower and upper energy levels, but $5p^65d^97s\ ^1D_2$ (108631 $cm^{-1}$) is from RW [11] while $5p^65d^97p\ ^3F_3$ (123348 $cm^{-1}$) is from M [25]. This indicates that the discrepancies between NIST and our calculations are due to the difference in the energy levels taken from various sources by NIST.

For shifted levels, McCann et al. [21] found average discrepancies between NIST energy levels measurements and $GRASP^0$ calculations of 8.6 % and 5.1 % for Au I and Au II, respectively. Given this, when comparing our experimental line wavelengths and intensities with theoretical results for those features where no NIST or other data are available (and hence their wavelengths remain unshifted), we search the calculations for wavelengths in a range of ± 8.6 % about the measured value for Au I, and ± 5.1 % for Au II. Hence for example for the unshifted Au I line we measured at 335.3 nm, we search the database of calculations for an appropriate match in the range 306.8–363.8 nm.

**Table 3:** Energy levels for Au I from the NIST database

| Level | T | Energy ($cm^{-1}$) | J | Ref |
|---|---|---|---|---|
| $5d^{10}6s$ | $^2S$ | 0.0000 | 1/2 | ED |
| $5d^96s^2$ | $^2D$ | 9161.1760 | 5/2 | ED |
| $5d^96s^2$ | $^2D$ | 21435.1843 | 3/2 | ED |
| $5d^{10}6p$ | $^2P$ | 37358.9884 | 1/2 | ED |
| $5d^{10}6p$ | $^2P$ | 41174.6091 | 3/2 | ED |
| $5d^96s6p$ | $^4P$ | 42163.5288 | 5/2 | ED |
| $5d^96s6p$ | $^4F$ | 45537.1826 | 7/2 | ED |
| $5d^96s6p$ | $^4F$ | 46174.9758 | 5/2 | ED |
| $5d^96s6p$ | $^4D$ | 46379.0433 | 5/2 | M |
| $5d^96s6p$ | $^4P$ | 47007.4320 | 3/2 | ED |
| $5d^96s6p$ | $^4F$ | 48697.1450 | 9/2 | ED |
| $5d^96s6p$ | $^4D$ | 51028.8870 | 7/2 | ED |
| $5d^96s6p$ | $^2D$ | 51231.5169 | 3/2 | ED |
| $5d^96s6p$ | $^4F$ | 51485.0101 | 3/2 | M |
| $5d^96s6p$ | $^2F$ | 51653.8958 | 5/2 | M |
| $5d^96s6p$ | $^2F$ | 52802.0772 | 7/2 | M |
| $5d^96s6p$ | $^4P$ | 53196.3084 | 1/2 | BG |
| $5d^96s6p$ | $^4D$ | 55732.5020 | 1/2 | M |
| $5d^96s6p$ | $^4D$ | 56105.7405 | 3/2 | ED |
| $5d^96s6p$ | $^4D$ | 58616.7604 | 5/2 | ED |
| $5d^96s6p$ | $^2P$ | 58845.4090 | 3/2 | ED |
| $5d^96s6p$ | $^2D$ | 59713.2226 | 5/2 | M |
| $5d^96s6p$ | $^2F$ | 61255.1414 | 5/2 | M |
| $5d^96s6p$ | $^2P$ | 61563.2837 | 3/2 | M |
| $5d^{10}6d$ | $^2D$ | 61951.6440 | 3/2 | M |
| $5d^{10}6d$ | $^2D$ | 62033.7275 | 5/2 | M |

**Note to Table 3:** The Ref column lists the source of the energy level data, where ED, BG and M are Ehrhardt & Davis [23], Brown & Ginter [24] and Moore [25], respectively.

**Table 4:** Energy levels for Au II from the NIST database

| Level | T | Energy ($cm^{-1}$) | J | Ref |
|---|---|---|---|---|
| $5p^6 5d^{10}$ | $^1S$ | 0.0000 | 0 | RW |
| $5p^6 5d^9 6s$ | $^3D$ | 15039.5738 | 3 | RW |
| $5p^6 5d^9 6s$ | $^3D$ | 17640.6112 | 2 | RW |
| $5p^6 5d^9 6s$ | $^3D$ | 27765.7536 | 1 | RW |
| $5p^6 5d^9 6s$ | $^1D$ | 29621.2467 | 2 | RW |
| $5p^6 5d^8 6s^2$ | $^3F$ | 40478.7430 | 4 | RW |
| $5p^6 5d^8 6s^2$ | $^1D$ | 48510.8878 | 2 | RW |
| $5p^6 5d^8 6s^2$ | $^3F$ | 52176.4977 | 3 | RW |
| $5p^6 5d^8 6s^2$ | $^3P$ | 58191.6271 | 2 | RW |
| $5p^6 5d^8 6s^2$ | $^3P$ | 58550.2157 | 0 | RW |
| $5p^6 5d^8 6s^2$ | $^1G$ | 61384.5217 | 4 | RW |
| $5p^6 5d^8 6s^2$ | $^3P$ | 61749.4092 | 1 | RW |
| $5p^6 5d^9 6p$ | $^3P$ | 63053.3078 | 2 | RW |
| $5p^6 5d^9 6p$ | $^3F$ | 65003.5810 | 3 | RW |
| $5p^6 5d^8 6s^2$ | $^1D$ | 68145.1075 | 2 | RW |
| $5p^6 5d^9 6p$ | $^3F$ | 72495.1160 | 4 | RW |
| $5p^6 5d^9 6p$ | $^1D$ | 73178.2746 | 2 | RW |
| $5p^6 5d^9 6p$ | $^3P$ | 73403.8287 | 1 | RW |
| $5p^6 5d^9 6p$ | $^3D$ | 74791.4678 | 3 | RW |
| $5p^65d^96p$ | $^3F$ | 76659.6905 | 2 | RW |
| $5p^65d^96p$ | $^3D$ | 81659.8157 | 1 | RW |
| $5p^65d^96p$ | $^3P$ | 82613.7730 | 0 | RW |
| $5p^65d^96p$ | $^1F$ | 85700.1894 | 3 | RW |
| $5p^65d^96p$ | $^1P$ | 85707.5638 | 1 | RW |
| $5p^65d^96p$ | $^3D$ | 86565.6546 | 2 | RW |
| $5p^65d^86s^2$ | $^1S$ | 91114.4412 | 0 | RW |
| $5p^65d^97s$ | $^3D$ | 108172.9399 | 3 | RW |
| $5p^65d^97s$ | $^1D$ | 108631.4224 | 2 | RW |
| $5p^65d^96d$ | $^3S$ | 116050.4869 | 1 | RW |
| $5p^65d^96d$ | $^3G$ | 116946.2724 | 4 | RW |
| $5p^65d^96d$ | $^3P$ | 117065.5569 | 2 | RW |
| $5p^65d^96d$ | $^1P$ | 117296.8831 | 1 | RW |
| $5p^65d^96d$ | $^3D$ | 117511.9682 | 3 | RW |
| $5p^65d^96d$ | $^1F$ | 117982.7412 | 3 | RW |
| $5p^65d^96d$ | $^1D$ | 118028.8309 | 2 | RW |
| $5p^65d^96d$ | $^3F$ | 118167.5389 | 4 | RW |
| $5p^65d^97p$ | $^3P$ | 119446.5271 | 2 | RW |
| $5p^65d^97p$ | $^1F$ | 120257.1565 | 3 | RW |
| $5p^65d^97s$ | $^3D$ | 120822.9620 | 1 | RW |
| $5p^65d^97s$ | $^3D$ | 121118.8138 | 2 | RW |
| $5p^65d^97p$ | $^3P$ | 121784.1511 | 2 | M |
| $5p^65d^97p$ | $^3P$ | 121861.9548 | 1 | M |
| $5p^65d^97p$ | $^3F$ | 123061.8224 | 4 | M |
| $5p^6 5d^9 7p$ | $^3F$ | 123343.7376 | 3 | M |

**Note to Table 4:** The Ref column lists the source of the energy level data, where RW and M are Rosberg & Wyart [11] and Moore [25], respectively.

There is specific Au I and Au II energy levels that generate very intense lines in our collisional-radiative plasma model. For Au I, the $5d^{10}6p$ $^2P_{3/2}$ (41175 $cm^{-1}$) energy level is involved in 3 transitions that produce the 479.41 nm, 481.3 nm and 506.6 nm lines which are very strong. The $5d^9 6s^2$ $^2D_{3/2}$ (21435 $cm^{-1}$), $5d^{10}6p$ $^2P_{1/2}$ (37359 $cm^{-1}$) and $5d^{10}6d$ $^2D_{3/2}$ (61952 $cm^{-1}$) levels are also in transitions that show intense lines. For Au II, the $5p^6 5d^8 6s^2$ $^1D_2$ (48511 $cm^{-1}$) level is in 3 strong transitions (401.72, 405.39 and 606.33 nm). The highest levels included in our calculations for Au I ($5d^{10}6s$ $^2S_{1/2}$) and Au II ($5p^6 5d^{10}$ $^1S_0$) are at 145204 $cm^{-1}$ and 362257 $cm^{-1}$, respectively.

Of the 48 experimentally observed lines of Au I listed in Table 5, 30 are matched with the NIST database i.e., ED [23], 2 with Bromley et al. [10] and 8 with PS [22]. Hence there are 8 lines that can only be matched to those generated in our plasma model, indicating that these are very likely to be new Au I identifications.

Table 6 contains a list of 40 well-resolved Au II lines we have identified within our experimental range. Only 6 of these are listed in the NIST database i.e., ED [23], 5 by Bromley et al. [10] and 16 in PS [22], leaving 13 possible new identifications.

Table 7 lists 33 lines in our Au spectra which cannot be matched to NIST, other references or transitions in our plasma model. Several of these, such as 490.24, 514.2 and 585.37 nm, are similar in intensity to some of the strong lines which are identified in the NIST database and ED [23]. These may be due to Au I transitions involving 7s and above levels, which unfortunately are not included in our atomic physics calculations. Some broad lines are also detected in our experiments, and listed in Table 7, which may be due to autoionising levels of Au I as suggested by ED [23]. The inclusion of additional energy levels (higher than 6s) and autoionising levels in the atomic physics calculations for Au I would be of interest to allow further study of these unidentified lines.

**Conclusions:**

We have reported a pilot study to investigate if a laser-produced plasma under high vacuum may be employed to identify lines of high Z materials such as Au relevant to NSMs. We find a total of 54 lines in our spectra, covering 350 – 1000 nm, that do not match with NIST or other references, nor arise from any known impurities of Au in our targets (Cu and Ag). In addition, we have generated synthetic spectra from collisional-radiative models of Au I and Au II using atomic data from GRASP$^0$ (General-purpose Relativistic Atomic Structure Package) and a parallelised version of the DARC (Dirac Atomic R-matrix Code) code. Of the 54 lines, 21 of the strongest were matched with those found in our Au I and Au II synthetic spectra, with good agreement between the theoretical and experimental line intensities. Given this, we tentatively identify these 21 lines as previously-unreported Au I and Au II transitions, and also list their associated energy levels. In addition, we detected 33 lines, including strong features such as 490.24, 514.2 and 585.37 nm, which are neither previously identified nor in our synthetic spectra. These may be Au I transitions involving the 7s and higher levels, which were not included in our GRASP$^0$ or DARC calculations. Several broad lines were also found which may be due to Au I autoionising levels, as suggested by ED [23]. Further work is now required to confirm the provisional identifications of the 21 lines, and also identify the 33 lines which remain unknown. This includes higher resolution experimental data, to better determine the line wavelengths. In addition, more extensive Au I and Au II atomic physics calculations are required, including both higher-lying and autoionising levels, to both help identify unknown transitions and reduce errors in the theoretical energy levels and hence wavelengths, to allow a better comparison with experiment. However, our pilot study does demonstrate that spectroscopy of laser-produced plasmas is well suited to identifying heavy element transitions that may not be detected using other experimental methods.

**Data Access Statement:**

The analysed data published in this paper are available from the corresponding author at s.chaurasia@qub.ac.uk on request. These, plus the raw data and atomic calculations on which the paper is based, have been stored on the Queen's University Research Data Management System, and once again are available from s.chaurasia@qub.ac.uk on request.

**Acknowledgements:**

This work was carried out using the laser facilities and equipment available in the Centre for Light-Matter Interactions (CLMI) at QUB. Part of the work reported here is presented in the thesis of M Charlwood [26]. We would like to thank Professor Steven Rose of Imperial College London for suggesting that laser-produced plasmas could be used for line identification work relevant to NSMs. Financial support for the project was provided by the UKRI Science and Technology Facilities Council via grant ST/T000198/1, while MC is grateful to the NI Department for the Economy (DfE) for the award of a PhD studentship.

**CRediT Authorship Contribution Statement:**

**M. Charlwood:** Investigation, Methodology, Analysis, Writing. **S. Chaurasia:** Conceptualization, Investigation, Methodology, Analysis, Writing. **M. McCann:** Simulations. **C. Ballance:** Simulations. **D. Riley:** Conceptualisation, Supervision, Review and editing, **F.P. Keenan:** Conceptualisation, Supervision, Writing, Review and editing.

**Table 5:** Experimental Au I lines and transitions

| Experimental $\lambda$ (nm) | NIST $\lambda$ (nm) | Sim $\lambda$ (nm) | Lower Level | $T_L$ | $J_L$ | Upper Level | $T_U$ | $J_U$ | Charge State | Note | Ref |
|---|---|---|---|---|---|---|---|---|---|---|---|
| 355.30 | 355.36 | 355.43 unshifted | $5p^6 5d^8 6s^2$ | $^3P$ | 2 | $5p^6 5d^9 6p$ | $^3D$ | 2 | Au II (Sim) Au I (NIST) | No NIST transition for Au I, matched with Au II | ED |
| 355.68 | 355.74 | 355.92 unshifted | $5d^8 6s^2 6p$ | $^2H$ | 11/2 | $5d^8 6s^2 6d$ | $^2I$ | 11/2 | Au I | No NIST transition, Charge states match | ED |
| 356.50 | 356.59 | 357.28 unshifted | $5p^6 5d^9 7s$ | $^1D$ | 2 | $5p^6 5d^8 6s6p$ | $^1D$ | 2 | Au II (Sim) Au I (NIST) | No NIST transition for Au I, matched with Au II | ED |
| 358.62 | 358.67 | 358.78 unshifted | $5d^8 6s^2 6p$ | $^4S$ | 11/2 | $5d^8 6s^2 6d$ | $^4H$ | 13/2 | Au I | No NIST transition, Charge states match | ED |
| 359.75 | 359.81 | 359.97 unshifted | $5d^86s^26p$ | $^4P$ | 5/2 | $5d^86s^26d$ | $^2G$ | 7/2 | Au I | No NIST transition, Charge states match | ED |
| 360.70 | | 360.81 unshifted | $5d^86s^26p$ | $^4D$ | 5/2 | $5d^96p^2$ | $^2D$ | 3/2 | Au I | | New |
| 361.98 | | 362.47 unshifted | $5d^86s^26d$ | $^2P$ | 3/2 | $5d^86s^26d$ | $^2F$ | 7/2 | Au I | | New |
| 365.02 | 365.07 | | $5d^{10}6p$ | $^2P$ | 1/2 | $5d^{10}8s$ | $^2S$ | 1/2 | Au I | | ED |
| 404.06 | 404.09 | 404.21 | $5d^96s^2$ | $^2D$ | 3/2 | $5d^96s6p$ | $^4F$ | 5/2 | Au I | Matching transitions and Charge states | ED |
| 404.56 | | | $5d^96s6p$ | $^4P$ | 3/2 | $5d^{10}11s$ | $^2S$ | 1/2 | Au I | 404.6nm | B |
| 406.46 | 406.51 | 406.62 | $5d^{10}6p$ | $^2P$ | 1/2 | $5d^{10}6d$ | $^2D$ | 3/2 | Au I | Matching transitions and Charge states | ED |
| 408.37 | 408.41 | 408.44 | $5d^96s6p$ | $^4F$ | 5/2 | $5d^96s(^3D_2)7s$ | $^4D$ | 3/2 | Au I | NIST Au I has 7s, Not in Sim | ED |
| | | | $5p^65d^86s^2$ | $^3F$ | 3 | $5p^65d^96p$ | $^3F$ | 2 | Au II<br>(Sim) | Au II close | |
| 408.94 | | 409.06<br>unshifted | $5d^86s^26p$ | $^2D$ | 3/2 | $5d^86s^26d$ | $^4P$ | 1/2 | Au I | | New |
| 410.23 | 410.17 | 410.31<br>unshifted | $5d^96s6p$<br>$5d^96s6p$ | $^4P$<br>$^2P$ | 3/2<br>1/2 | $5d^{10}9d$<br>$5d^96p^2$ | $^2D$<br>$^2P$ | 5/2<br>1/2 | Au I | NIST Au I has 9d, Not in Sim<br>Another<br>Au I<br>close | ED |
| 424.21 | 424.18 | 424.30<br>unshifted | $5d^{10}6p$<br>$5d^86s^26p$ | $^2P$<br>$^2F$ | 3/2<br>7/2 | $5d^{10}8s$<br>$5d^96p^2$ | $^2S$<br>$^2G$ | 1/2<br>7/2 | Au I<br>Au I<br>(Sim) | NIST Au I has 8s, Not in Sim Another<br>Au I close | ED |
| 431.48 | 431.51 | 431.72<br>unshifted | $5d^96s6p$<br>$5d^86s6p^2$ | $^4F$<br>$^4D$ | 7/2<br>5/2 | $5d^96s(^3D_37s$<br>$5d^86s^26d$ | $^4D$<br>$^4F$ | 5/2<br>5/2 | Au I<br>Au I<br>(Sim) | NIST Au I has 7s, Not in Sim Another<br>Au I close | ED |
| 443.75 | 443.73 | | $5d^9 6s6p$ | $^4F$ | 5/2 | $5d^9 6s(^3D_3)7s$ | $^4D$ | 5/2 | Au I | | ED |
| 448.81 | 448.83 | | $5d^9 6s6p$ | $^4F$ | 7/2 | $5d^9 6s(^3D_3)7s$ | $^4D$ | 7/2 | Au I | | ED |
| 460.77 | 460.75 | | $5d^9 6s6p$ | $^4P$ | 3/2 | $5d^9 6s(^3D_3)7s$ | $^4D$ | 5/2 | Au I | | ED |
| 479.29 | 479.26 | 479.41 | $5d^{10} 6p$ | $^2P$ | 3/2 | $5d^{10} 6d$ | $^2D$ | 5/2 | Au I | Matching transitions and Charge states | ED |
| 481.16 | 481.16 | 481.3 | $5d^{10} 6p$ | $^2P$ | 3/2 | $5d^{10} 6d$ | $^2D$ | 3/2 | Au I | Matching transitions and Charge states | ED |
| 490.24 | | | | | | | | | Au I | 490.23nm | PS |
| 506.43 | 506.46 | 506.6 | $5d^9 6s^2$ | $^2D$ | 3/2 | $5d^{10} 6p$ | $^2P$ | 3/2 | Au I | Matching transitions and Charge states | ED |
| 514.74 | 514.74 | | $5d^9 6s6p$ | $^2D$ | 3/2 | $5d^9 6s(^3D_2)7s$ | $^4D$ | 3/2 | Au I | | ED |
| 516.74 | | 516.87 unshifted | $5d^8 6s^2 6p$ | $^4G$ | 7/2 | $5d^9 6p^2$ | $^2D$ | 5/2 | Au I | Low Intensity, Sim<Exp | New |
| 517.86 | | 517.49 unshifted | $5d^9 6s6p$ | $^2D$ | 3/2 | $5d^9 6p^2$ | $^4D$ | 1/2 | Au I | Low Intensity, Sim<Exp | New |
| 523.02 | 523.03 | | $5d^96s6p$ | $^4F$ | 9/2 | $5d^96s(^3D_3)7s$ | $^4D$ | 7/2 | Au I | | ED |
| 526.18 | 526.18 | | $5d^96s6p$ | $^2D$ | 5/2 | $5d^96s(^3D_2)7s$ | $^4D$ | 3/2 | Au I | | ED |
| 538.42 | | 538.66<br>unshifted | $5d^86s^26p$ | $^2D$ | 5/2 | $5d^86s^26d$ | $^4G$ | 7/2 | Au I | Weak | New |
| 553.29 | | | | | | | | | Au I | 553.2 nm | PS |
| 564.14 | | | | | | | | | Au I | 564.13nm | PS |
| 565.55 | 565.58 | | $5d^96s6p$ | $^4D$ | 7/2 | $5d^96s(^3D_3)7s$ | $^4D$ | 5/2 | Au I | | ED |
| 572.13 | 572.14 | 572.24 | $5d^96s6p$<br>$5p^65d^97s$ | $^2D$<br>$^3D$ | 3/2<br>3 | $5d^96s(^3D_3)7s$<br>$5p^65d^86s6p$ | $^4D$<br>$^3F$ | 5/2<br>2 | Au I<br>Au II<br>(Sim) | NIST Au I has 7s,<br>Not in Sim<br>Au II close | ED |
| 572.67 | | | | | | | | | Au I | 572.67nm | PS |
| 574.18 | | | | | | | | | Au I | 574.12nm | PS |
| 583.73 | 583.74 | | $5d^{10}6p$ | $^2P$ | 1/2 | $5d^{10}7s$ | $^2S$ | 1/2 | Au I | | ED |
| 586.30 | 586.29 | | $5d^96s6p$ | $^2D$ | 5/2 | $5d^96s(^3D_3)7s$ | $^4D$ | 5/2 | Au I | | ED |
| 590.12 | | | $5d^96s6p$ | $^4D$ | 3/2 | $5d^{10}12d$ | $^2D$ | 5/2 | Au I | 589.82nm | B |
| 595.73 | 595.70 | | $5d^96s6p$ | $^4D$ | 7/2 | $5d^96s(^3D_3)7s$ | $^4D$ | 7/2 | Au I | | ED |
| 596.29 | 596.27 | | | | | | | | Au I | | ED |
| 627.78 | 627.82 | 628 | $5d^96s^2$ | $^2D$ | 3/2 | $5d^{10}6p$ | $^2P$ | 1/2 | Au I | Matching transitions and Charge states | ED |
| 630.12 | | | | | | | | | Au I | 630.45nm | PS |
| 633.97 | | 633.85 | $5d^96s6p$ | $^4F$ | 5/2 | $5d^{10}6d$ | $^2D$ | 3/2 | Au I | 633.67nm | PS |
| 638.71 | | 638.79 | $5d^86s^26p$ | $^4D$ | 3/2 | $5d^86s6p^2$ | $^2D$ | 5/2 | Au I | 638.77nm | PS |
| 642.79 | | 642.16 | $5d^96s6p$ | $^4D$ | 5/2 | $5d^{10}6d$ | $^2D$ | 3/2 | Au I | Broad | New |
| 649.94 | | 649.39 | $5d^86s^26p$ | $^4G$ | 11/2 | $5d^96p^2$ | $^4F$ | 9/2 | Au I | | New |
| | | unshifted | | | | | | | | | |
| 665.24 | 665.29 | 665.5 | $5d^96s6p$ | $^4P$ | 3/2 | $5d^{10}6d$ | $^2D$ | 5/2 | Au I | No NIST transition, Charge states match | ED |
| 751.05 | 751.07 | | $5d^{10}6p$ | $^2P$ | 3/2 | $5d^{10}7s$ | $^2S$ | 1/2 | Au I | | ED |

**Notes to Table 5:** The NIST column contains the wavelength from the NIST database where available. Sim is the theoretical wavelength from our atomic physics calculations, with 'unshifted' indicating that the theoretical energy levels for the line could not be shifted to those in NIST as they are absent from the latter. No NIST transition indicates that NIST lists a wavelength for the line but does not identify the transition. Ref lists the source of the NIST data, with PS, ED and B denoting Platt & Sawyer [22], Ehrhardt & Davis [23] and Bromley et al. [10], respectively. Where there is no NIST listing, we provide a Bromley et al. [10] wavelength, if available.

**Table 6:** Experimental Au II lines and transitions

| Experimental $\lambda$ (nm) | NIST $\lambda$ (nm) | Sim $\lambda$ (nm) | Lower Level | $T_L$ | $J_L$ | Upper Level | $T_U$ | $J_U$ | Charge State | Note | Ref |
|---|---|---|---|---|---|---|---|---|---|---|---|
| 363.26 | 363.32 | 363.43 | $5p^65d^86s^2$ | $^3P$ | 2 | $5p^65d^96p$ | $^1P$ | 1 | Au II | No NIST transition, Charge states match | ED |
| 401.55 | 401.61 | 401.72 | $5p^65d^86s^2$ | $^1D$ | 2 | $5p^65d^96p$ | $^3P$ | 1 | Au II | No NIST transition, Charge states match | ED |
| 405.22 | 405.28 | 405.39 | $5p^65d^86s^2$ | $^1D$ | 2 | $5p^65d^96p$ | $^1D$ | 2 | Au II | No NIST transition, Charge states match | ED |
| 407.58 | 407.64 | 407.75 | $5p^6 5d^8 6s^2$ | $^3F$ | 4 | $5p^6 5d^9 6p$ | $^3F$ | 3 | Au II | No NIST transition, Charge states match | ED |
| 417.23 | | 417.39 unshifted | $5p^6 5d^8 6s^2$ | $^3P$ | 1 | $5p^6 5d^9 6p$ | $^1P$ | 1 | Au II | | New |
| 435.78 | | 436.23 | $5p^6 5d^9 6p$ | $^1P$ | 1 | $5p^6 5d^9 7s$ | $^1D$ | 2 | Au II | 435.95nm | PS |
| 442.08 | 442.06 | 442.18 | $5p^6 5d^8 6s^2$ | $^3F$ | 3 | $5p^6 5d^9 6p$ | $^3D$ | 3 | Au II | No NIST transition, Charge states match | ED |
| 476.03 | 476.02 | 476.15 | $5d^9 6s6p$ | $^3F$ | 5/2 | $5d^9 6s(^3D_3)7s$ | $^1D$ | 7/2 | Au II | No NIST transition, Charge states match | ED |
| 492.06 | | 492.16 | $5p^5 5d^8 6s^2$ | $^3F$ | 3 | $5p^6 5d^9 6p$ | $^3F$ | 4 | Au II | 492.06nm | B |
| 501.3 | 501.29 | | $5d^96d$ | | 1 | $5d^97p$ | | 2 | Au II | 501.29nm | B |
| 502.05 | | 502.25 | $5p^65d^86s^2$ | $^3P$ | 1 | $5p^65d^96p$ | $^3D$ | 1 | Au II | | New |
| 541.33 | | 541.48 | $5p^65d^86s^2$ | $^3P$ | 2 | $5p^65d^96p$ | $^3F$ | 2 | Au II | 541.33nm | PS |
| 542.8 | | 542.87 | $5p^65d^86s^2$ | $^1D$ | 2 | $5p^65d^96p$ | $^3D$ | 2 | Au II | 542.72nm | PS |
| 552.03 | | 552.22<br>unshifted | $5p^65d^86s6p$ | $^5D$ | 3 | $5p^65d^97s$ | $^3D$ | 3 | Au II | 552.02nm | PS |
| 564.57 | | 564.63<br>unshifted | $5p^65d^96p$ | $^3P$ | 1 | $5p^65d^86s^2$ | $^1S$ | 0 | Au II | 564.49nm | PS |
| 578.29 | | | $5d^96d$ | | 1 | $5d^97p$ | | 1 | Au II | 578.09nm | B |
| 602.25 | | 602.42 | $5p^65d^86s^2$ | $^3P$ | 2 | $5p^65d^96p$ | $^3D$ | 3 | Au II | | New |

| | | | | | | | | | | |
|---|---|---|---|---|---|---|---|---|---|---|
| 606.16 | 606.33 | $5p^65d^86s^2$ | $^1D$ | 2 | $5p^65d^96p$ | $^3F$ | 3 | Au II | 606.19nm | PS |
| 621.26 | 621.13 unshifted | $5p^65d^97s$ | | 2 | $5p^65d^86s6p$ | | 3 | Au II | | New |
| 641.61 | 641.29 unshifted | $5p^65d^97s$ | $^3D$ | 3 | $5p^65d^86s6p$ | $^3F$ | 3 | Au II | | New |
| 658.84 | 659.16 | $5p^65d^97s$ | $^3D$ | 3 | $5p^65d^97p$ | $^3F$ | 3 | Au II | 658.94nm | PS |
| 660.42 | | $5d^86s6p$ | | 3 | $5d^96d$ | | 2 | Au II | 660.47nm | B |
| 671.31 | 671.64 | $5p^65d^97s$ | $^3D$ | 3 | $5p^65d^97p$ | $^3F$ | 4 | Au II | 671.43nm | PS |
| 673.03 | 673.24 | $5p^65d^86s^2$ | $^3P$ | 0 | $5p^65d^96p$ | $^3P$ | 1 | Au II | | New |
| 679.29 | 679.7 | $5p^65d^97s$ | $^1D$ | 2 | $5p^65d^97p$ | $^3F$ | 3 | Au II | 679.48nm | PS |
| 682.65 | 682.88 unshifted | $5p^65d^86s6p$ | $^5D$ | 2 | $5p^65d^97d$ | $^3S$ | 1 | Au II | Broad | New |
| 685.81 | 685.79 | $5p^65d^96s$ | $^3D$ | 3 | $5p^65d^96s$ | $^1D$ | 2 | Au II | | New |
| 687.42 | 687.64 | $5p^65d^86s^2$ | $^1D$ | 2 | $5p^65d^96p$ | $^3P$ | 2 | Au II | | New |
| 733.4 | | $5d^86s6p$ | | 2 | $5d^96d$ | | 1 | Au II | 733.4nm | B |
| 734.36 | 734.69 | $5p^65d^97s$ | $^3D$ | 3 | $5p^65d^97p$ | $^3P$ | 2 | Au II | 734.44nm | PS |
| 746.05 | | | | | | | | Au II | 745.71nm | PS |
| 755.52 | 755.83 | $5p^65d^97s$ | $^1D$ | 2 | $5p^65d^97p$ | $^3P$ | 1 | Au II | 755.58nm | PS |
| 759.93 | 760.30 | $5p^65d^97s$ | $^1D$ | 2 | $5p^65d^97p$ | $^3P$ | 2 | Au II | 760.04nm | PS |
| 768.90 | 769.42 unshifted | $5p^65d^97p$ | $^3P$ | 1 | $5p^65d^97d$ | $^1P$ | 1 | Au II | | New |
| 779.30 | 779.6 | $5p^65d^86s^2$ | $^3F$ | 3 | $5p^65d^96p$ | $^3F$ | 3 | Au II | | New |
| 827.23 | 827.53 | $5p^65d^97s$ | $^3D$ | 3 | $5p^65d^97p$ | $^1F$ | 3 | Au II | 827.29nm | PS |
| 859.87 | 860.16 | $5p^65d^97s$ | $^1D$ | 2 | $5p^65d^97p$ | $^1F$ | 3 | Au II | 859.91nm | PS |
| 868.86 | 868.86 unshifted | $5p^65d^97p$ | $^3F$ | 4 | $5p^65d^97d$ | $^3G$ | 5 | Au II | | New |
| 886.68 | 887.03 | $5p^65d^97s$ | $^3D$ | 3 | $5p^65d^97p$ | $^3P$ | 2 | Au II | 886.76nm | PS |
| 919.05 | 919.39 | $5p^65d^86s^2$ | $^3F$ | 2 | $5p^65d^96p$ | $^3P$ | 2 | Au II | | New |

**Notes to Table 6:** The NIST column contains the wavelength from the NIST database where available. Sim is the theoretical wavelength from our atomic physics calculations, with 'unshifted' indicating that the theoretical energy levels for the line could not be shifted to those in NIST as they are absent from the latter. No NIST transition indicates that NIST lists a wavelength for the line but does not identify the transition. Ref lists the source of the NIST data, with PS, ED and B denoting Platt & Sawyer [22], Ehrhardt & Davis [23] and Bromley et al. [10], respectively. Where there is no NIST listing, we provide a Bromley et al. [10] wavelength, if available.

**Table 7**: Experimental Au Lines that remain unidentified

| Experimental $\lambda$ (nm) | Note |
|---|---|
| 458.83 | Broad |
| 514.20 | Sharp |
| 536.46 | Broad |
| 538.09 | Broad |
| 539.5 | Broad |
| 557.84 | Broad |
| 558.89 | Broad |
| 559.87 | Broad |
| 571.09 | Broad |
| 575.97 | Broad |
| 578.84 | Broad |
| 581.87 | Small |
| 585.37 | Sharp |
| 588.20 | Very Broad |
| 591.98 | Broad |
| 593.30 | Clear Peak |
| 607.26 | Broad |
| 631.07 | Broad |
| 643.41 | Broad |
| 644.21 | Broad |
| 652.30 | Clear Peak |
| 654.12 | Broad |
| 670.80 | Clear Peak |
| 671.96 | Broad |
| 690.16 | Broad |
| 724.04 | Broad |
| 726.04 | |
| 726.43 | |
| 727.98 | Broad |
| 729.77 | Broad |
| 737.12 | |

| |
|---|
| 756.77 |
| 920.25 |

## Annexure A

**Table 2.1**: List of Au, Ag, Pt, Pd and Cu observed Lines that were matched to NIST to produce Figure 2.

| Au Observed (nm) | Au NIST (nm) | Ag Observed (nm) | Ag NIST (nm) | Pt Observed (nm) | Pt NIST (nm) | Pd Observed (nm) | Pd NIST (nm) | Cu Observed (nm) | Cu NIST (nm) |
|---|---|---|---|---|---|---|---|---|---|
| 352.31 | 352.33 | 328.08 | 328.07 | 340.81 | 340.81 | 330.24 | 330.21 | 303.48 | 303.46 |
| 358.66 | 358.67 | 338.29 | 338.29 | 342.80 | 342.79 | 337.28 | 337.30 | 327.42 | 327.40 |
| 363.30 | 363.32 | 346.91 | 346.92 | 343.17 | 343.19 | 340.45 | 340.46 | 330.84 | 330.84 |
| 380.41 | 380.40 | 368.34 | 368.33 | 344.77 | 344.78 | 342.11 | 342.12 | 332.16 | 332.16 |
| 387.45 | 387.47 | 392.04 | 392.01 | 345.38 | 345.38 | 343.34 | 343.34 | 345.02 | 345.03 |
| 388.02 | 388.03 | 398.51 | 398.52 | 345.57 | 345.58 | 344.12 | 344.14 | 351.16 | 351.18 |
| 389.22 | 389.23 | 405.51 | 405.55 | 347.76 | 347.77 | 346.04 | 346.07 | 359.88 | 359.91 |
| 389.81 | 389.79 | 421.08 | 421.10 | 348.32 | 348.34 | 348.09 | 348.12 | 360.16 | 360.19 |
| 392.76 | 392.77 | 438.51 | 438.51 | 348.51 | 348.53 | 348.95 | 348.98 | 362.05 | 362.04 |
| 395.92 | 395.91 | 447.63 | 447.60 | 350.55 | 350.54 | 351.67 | 351.69 | 362.74 | 362.73 |
| 401.57 | 401.61 | 462.01 | 462.00 | 351.43 | 351.44 | 355.28 | 355.31 | 365.15 | 365.18 |
| 405.24 | 405.28 | 466.84 | 466.85 | 353.56 | 353.59 | 357.09 | 357.11 | 367.65 | 367.69 |
| 408.38 | 408.41 | 478.84 | 478.84 | 354.82 | 354.85 | 360.93 | 360.95 | 368.67 | 368.66 |
| 424.20 | 424.18 | 502.72 | 502.73 | 355.11 | 355.14 | 363.45 | 363.47 | 369.32 | 369.33 |
| 431.52 | 431.51 | 514.28 | 514.28 | 357.20 | 357.20 | 369.05 | 369.03 | 369.50 | 369.54 |
| 442.08 | 442.06 | 520.88 | 520.91 | 357.72 | 357.72 | 371.90 | 371.89 | 380.50 | 380.52 |
| 443.75 | 443.73 | 531.70 | 531.72 | 358.72 | 358.73 | 379.95 | 379.92 | 385.09 | 385.11 |
| 448.84 | 448.83 | 537.40 | 537.38 | 360.51 | 360.48 | 383.24 | 383.23 | 396.42 | 396.42 |
| 460.74 | 460.75 | 544.10 | 544.09 | 360.78 | 360.79 | 389.43 | 389.42 | 402.26 | 402.26 |
| 475.99 | 476.02 | 546.56 | 546.55 | 362.81 | 362.81 | 395.85 | 395.86 | 404.28 | 404.25 |
| 479.28 | 479.26 | 547.16 | 547.16 | 363.87 | 363.88 | 416.96 | 416.98 | 406.25 | 406.26 |
| 481.17 | 481.16 | 554.34 | 554.32 | 364.30 | 364.32 | 421.27 | 421.29 | 420.17 | 420.17 |
| 506.46 | 506.46 | 555.84 | 555.81 | 366.42 | 366.39 | 447.36 | 447.36 | 422.78 | 422.79 |
| 514.74 | 514.74 | 573.40 | 573.42 | 366.84 | 366.84 | 478.84 | 478.82 | 427.52 | 427.51 |
| 523.03 | 523.03 | 608.36 | 608.38 | 367.20 | 367.20 | 481.75 | 481.75 | 437.84 | 437.84 |
| 526.18 | 526.18 | 633.10 | 633.13 | 367.27 | 367.28 | 487.56 | 487.54 | 450.60 | 450.60 |
| 565.58 | 565.58 | 640.38 | 640.36 | 367.55 | 367.55 | 511.67 | 511.70 | 450.95 | 450.94 |
| 572.17 | 572.14 | 650.56 | 650.57 | 373.05 | 373.01 | 523.45 | 523.49 | 453.11 | 453.08 |
| 583.75 | 583.74 | 669.70 | 669.72 | 386.82 | 386.84 | 529.54 | 529.56 | 453.99 | 453.97 |
| 586.30 | 586.29 | 693.97 | 694.00 | 387.59 | 387.57 | 531.22 | 531.26 | 455.57 | 455.59 |
| 595.73 | 595.70 | 699.92 | 699.89 | 390.10 | 390.07 | 534.53 | 534.51 | 457.55 | 457.57 |
| 596.29 | 596.27 | 768.78 | 768.78 | 390.47 | 390.44 | 554.28 | 554.28 | 458.68 | 458.70 |
| 627.79 | 627.82 | 825.48 | 825.48 | 394.87 | 394.86 | 554.70 | 554.70 | 465.11 | 465.11 |
| 656.28 | 656.27 | 826.27 | 826.29 | 396.64 | 396.64 | 561.95 | 561.95 | 467.50 | 467.47 |
| 665.27 | 665.29 | 827.36 | 827.35 | 397.02 | 397.01 | 564.28 | 564.27 | 468.21 | 468.20 |
| 751.09 | 751.07 | 870.50 | 870.54 | 402.40 | 402.38 | 566.97 | 567.01 | 468.80 | 468.78 |
| | | | | 403.42 | 403.42 | 568.98 | 569.01 | 469.74 | 469.75 |
| | | | | 404.64 | 404.65 | 569.47 | 569.51 | 470.44 | 470.46 |
| | | | | 406.15 | 406.17 | 573.63 | 573.66 | 476.77 | 476.75 |
| | | | | 406.61 | 406.61 | 678.41 | 678.45 | 483.23 | 483.22 |
| | | | | 407.18 | 407.21 | 683.35 | 683.34 | 488.98 | 488.97 |
| | | | | 408.14 | 408.15 | 701.66 | 701.64 | 493.74 | 493.75 |
| | | | | 408.51 | 408.48 | 730.98 | 731.01 | 501.29 | 501.26 |
| | | | | 409.21 | 409.23 | 736.79 | 736.81 | 501.69 | 501.66 |

| | | | | | |
|---|---|---|---|---|---|
| 410.53 | 410.55 | 739.18 | 739.19 | 502.16 | 502.13 |
| 411.87 | 411.87 | 748.68 | 748.69 | 503.45 | 503.44 |
| 414.84 | 414.83 | 776.36 | 776.40 | 504.16 | 504.13 |
| 416.44 | 416.45 | 778.63 | 778.66 | 505.21 | 505.18 |
| 419.24 | 419.24 | 791.56 | 791.58 | 505.95 | 505.94 |
| 420.16 | 420.12 | 813.27 | 813.28 | 506.05 | 506.06 |
| 422.42 | 422.38 | 835.33 | 835.35 | 506.72 | 506.71 |
| 426.07 | 426.10 | 853.27 | 853.27 | 507.60 | 507.62 |
| 427.15 | 427.12 | 859.93 | 859.91 | 508.41 | 508.40 |
| 427.42 | 427.39 | | | 508.91 | 508.89 |
| 428.86 | 428.84 | | | 510.53 | 510.55 |
| 435.83 | 435.83 | | | 511.17 | 511.19 |

| | | | |
|---|---|---|---|
| 439.22 | 439.18 | 512.43 | 512.45 |
| 441.45 | 441.43 | 514.42 | 514.41 |
| 444.29 | 444.25 | 515.29 | 515.32 |
| 449.89 | 449.89 | 515.83 | 515.84 |
| 451.15 | 451.12 | 521.80 | 521.82 |
| 451.43 | 451.41 | 521.99 | 522.01 |
| 452.11 | 452.09 | 526.98 | 527.00 |
| 452.32 | 452.30 | 529.25 | 529.25 |
| 454.80 | 454.79 | 531.64 | 531.60 |
| 455.27 | 455.24 | 532.44 | 532.44 |
| 458.10 | 458.07 | 533.82 | 533.85 |
| 464.11 | 464.08 | 535.66 | 535.68 |
| 468.44 | 468.41 | 536.95 | 536.98 |
| 473.76 | 473.76 | 539.00 | 539.00 |
| 500.29 | 500.26 | 541.86 | 541.85 |
| 503.88 | 503.85 | 546.34 | 546.31 |
| 504.41 | 504.41 | 555.51 | 555.49 |
| 505.50 | 505.53 | 562.14 | 562.17 |
| 505.97 | 505.95 | 565.20 | 565.17 |
| 508.27 | 508.23 | 570.00 | 570.02 |
| 519.39 | 519.39 | 572.15 | 572.18 |
| 519.91 | 519.92 | 576.14 | 576.12 |
| 522.33 | 522.31 | 578.19 | 578.21 |
| 522.77 | 522.76 | 583.35 | 583.35 |
| 526.12 | 526.08 | 584.25 | 584.25 |
| 530.14 | 530.10 | 585.07 | 585.07 |
| 576.39 | 576.36 | 590.12 | 590.12 |
| 584.03 | 584.01 | 594.17 | 594.18 |
| 584.50 | 584.48 | 596.55 | 596.53 |
| 586.10 | 586.08 | 600.03 | 600.01 |
| 602.63 | 602.60 | 608.04 | 608.03 |
| 628.39 | 628.35 | 615.43 | 615.42 |
| 631.84 | 631.84 | 618.89 | 618.87 |
| 632.66 | 632.66 | 620.87 | 620.90 |
| 664.82 | 664.83 | 621.70 | 621.69 |
| 671.05 | 671.04 | 621.98 | 621.98 |
| 675.98 | 676.00 | 626.19 | 626.18 |
| 683.80 | 683.81 | 626.82 | 626.83 |
| 684.25 | 684.26 | 627.31 | 627.33 |
| 709.50 | 709.48 | 630.09 | 630.10 |
| 711.39 | 711.37 | 631.12 | 631.13 |
| 721.76 | 721.76 | 635.77 | 635.74 |
| 762.61 | 762.62 | 637.72 | 637.72 |
| 783.04 | 783.04 | 642.38 | 642.39 |
| 820.48 | 820.45 | 644.16 | 644.17 |
| 822.47 | 822.48 | 644.88 | 644.86 |
| | | 647.43 | 647.42 |
| | | 648.47 | 648.44 |

| | |
|---|---|
| 650.85 | 650.84 |
| 651.27 | 651.25 |
| 662.18 | 662.16 |
| 662.43 | 662.43 |
| 666.06 | 666.10 |
| 669.31 | 669.27 |
| 674.14 | 674.14 |
| 677.94 | 677.94 |
| 687.20 | 687.22 |
| 690.58 | 690.59 |
| 691.55 | 691.57 |
| 696.92 | 696.88 |
| 743.79 | 743.82 |
| 757.05 | 757.01 |
| 766.47 | 766.46 |
| 772.66 | 772.66 |
| 793.29 | 793.31 |
| 809.25 | 809.26 |
| 847.71 | 847.73 |
| 951.58 | 951.56 |
| 953.03 | 953.03 |
| 966.38 | 966.41 |
| 973.18 | 973.21 |
| 1017.66 | 1017.64 |